\documentclass[prl,reprint,aps,showpacs]{revtex4-2}

\usepackage{amsmath,amssymb}
\usepackage{bm}
\usepackage{graphicx}

\begin{document}

\title{Nonreciprocal Quantum Mpemba Effect}

\author{Wei-Bin Yan,$^{1}$ Ying-Jie Zhang,$^{1,*}$\thanks{yingjiezhang@qfnu.edu.cn} Yun-Jie Xia,$^{1}$ Heng Fan,$^{2,3,4,5,\ddagger}$\thanks{hfan@iphy.ac.cn} and Zhong-Xiao Man$^{1,\dagger}$\thanks{manzhongxiao@163.com}}

\affiliation{$^{1}$College of Physics and Engineering, Qufu Normal University, Qufu 273165, China}

\affiliation{$^{2}$Beijing National Laboratory for Condensed Matter Physics, Institute of Physics, Chinese Academy of Sciences, Beijing 100190, China}

\affiliation{$^{3}$School of Physical Sciences, University of Chinese Academy of Sciences, Beijing 100049, China}

\affiliation{$^{4}$Hefei National Laboratory, Hefei 230088, China}

\affiliation{$^{5}$Beijing Key Laboratory of Advanced Quantum Technology, Beijing 100190, China }

\begin{abstract}
We demonstrate a nonreciprocal quantum Mpemba effect. Consider a broad class of open quantum systems, each coupled to two isomorphic reservoirs through symmetric ports. Interchanging the parameters of the two reservoirs---a discrete operation we call the swap---turns the quantum Mpemba effect on or off without changing the initial states. The swap modifies the Liouvillian, yet a structural symmetry pins the eigenvalues while rotating only the eigenvectors. The nonreciprocity therefore leaves no trace in the spectrum and is carried entirely by the eigenvectors. Concretely, the swap alters the far state's projection onto the slowest mode, switching whether it bypasses the slowest relaxation channel. At a Liouvillian exceptional point, the far state's relaxation switches from bypassing the slowest mode to avoiding the critical slowing, with the on--off contrast intact. There the spectrum-independent mechanism takes its purest form.
\end{abstract}

\maketitle

The Mpemba effect is a counterintuitive nonequilibrium phenomenon in which a state prepared farther from equilibrium can relax to it faster than one prepared closer. First reported in the cooling of water \cite{Mpemba1969PhysEduc}, it has since been documented in classical systems \cite{Lu2017PNAS,I2019PRX,A2017PRL,A2020Nature,G2026PhysRep}. Quantum realizations of the Mpemba effect reveal physics beyond classical energy-landscape mechanisms, giving rise to the quantum Mpemba effect (QME)~\cite{F2025NatRevPhys,G2026PhysRep}, studied in both isolated~\cite{F2023NatCommun,C2024PRL,A2025JPhysA,S2024PRL,Murciano2024JSTAT,A2025PRB2,Yamashika2024PRB,L2024PRL,Y2026PRL,T2025PRB} and open settings~\cite{S2024PRL1,A2023PRL,A2024PRL,X2024PRR,D2025PRL,F2023PRE,I2026arXiv,T2025PRL,P2025arXiv,J2025NatCommun,Y2025arXiv,M2026NJP,S2022PRA,R2026PRL,G2026arXiv,Longhi2025Quantum,Qian2025PRB,Westhoff2025PRA,R2025PRL,M2024PRL,D2024PRA,I2025PRL,Lejeune2026JPhysA,Zhou2023PRR,Chatterjee2024PRA,Z-Z2026PRL}. In open systems, the non-Hermitian Liouvillian dynamics provides an effective description of anomalous relaxation. For fixed dynamics, whether a state shows anomalous acceleration depends on its overlap with the slow-decaying mode. A vanishing overlap decouples the state from the slowest mode, bypassing it entirely and producing an exponential speedup~\cite{Carollo2021PRL}---the strong Mpemba effect. Faster relaxation directly benefits quantum state preparation and thermodynamic cycles.

Nonreciprocity---the breaking of reciprocity under role exchange---is a far broader and well-established principle. Appearing across many fields and in many forms \cite{M2026arXiv}, it has also been studied extensively in open and driven-dissipative systems. For instance, it gives rise to Liouvillian skin effects~\cite{T2021PRL,X2023PRB}, dissipation-induced nonreciprocal photon blockade~\cite{Li2024npjQI}, nonreciprocal quantum batteries~\cite{Ahmadi2024PRL}, interaction-mediated directional drift~\cite{P2026arXiv}, nonreciprocal synchronization~\cite{N2025PRX},  and a nonreciprocal response to temperature differences in a mesoscopic electronic conductor~\cite{Balduque2025PRL}. Yet, to our knowledge, none of these advances has made the QME itself nonreciprocal. This raises a sharp question: can the QME itself be nonreciprocal?

We answer this question by introducing a nonreciprocal quantum Mpemba effect (NQME). We consider a broad class of open quantum systems coupled to two isomorphic reservoirs through two symmetric ports. The parameters of the reservoirs differ by only a small bias. A swap operation interchanges the two reservoirs' roles, realized by interchanging their parameters. Now fix two initial states, one far from the steady state and one near it. Before the swap, the far state relaxes faster than the near one and reaches the steady state first. After the swap, the same far state no longer overtakes the near one. The same pair of states thus shows the Mpemba effect before the swap but not after. This is a property of the chosen states, not of the system alone, and it makes the quantum Mpemba effect nonreciprocal.

The mechanism behind the NQME is clean, revealing, and goes deeper than the nonreciprocity alone. The swap leaves any Liouvillian eigenvalue unchanged. An eigenvalue sets the decay rate of its eigenmode, a scalar. The swap fixes all of them. An eigenvector is a direction, and the swap rotates it. The nonreciprocity therefore cannot come from the eigenvalues. The eigenvectors, by contrast, carry the entire effect. Concretely, the far state starts orthogonal to the slowest eigenvector, so it skips the slow relaxation and reaches the steady state first. The swap rotates that eigenvector, the far state loses its orthogonality, and the speedup is gone.

A Liouvillian exceptional point (EP)~\cite{Minganti2019PRA} strips the mechanism to its essentials, yet the NQME persists. When eigenvectors coalesce and share the same decay, any spectral distinction vanishes. Before the swap, the far state relaxes as a clean exponential. After the swap, the same state is caught by a critical slowing, a non-exponential decay with an added time factor. It is a nonreciprocal critical slowing. With only one eigenvalue, the nonreciprocity takes its purest spectrum-independent form.

\begin{figure}[t]
\centering
\includegraphics[width=\columnwidth]{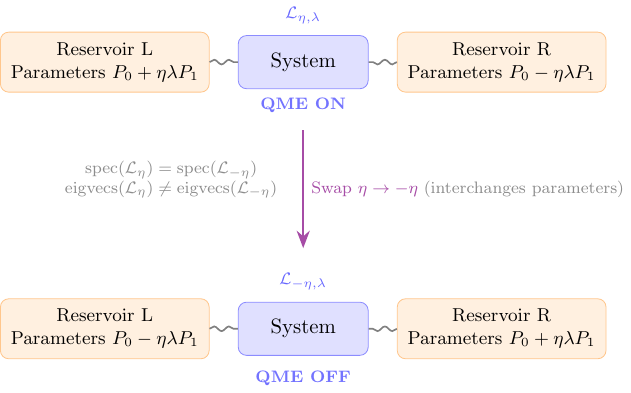}
\caption{Schematic of the nonreciprocal quantum Mpemba effect. Two reservoirs with parameters $P_0\pm\eta\lambda P_1$ couple to the system. The swap $\eta\to-\eta$ interchanges their parameters, turning the quantum Mpemba effect on and off for the same initial state.}
\label{fig:1}
\end{figure}

\textit{Model}. -- We consider the system Hilbert space $\mathcal H$ with $\dim\mathcal H=d<\infty$ and the Liouville space $\mathcal B(\mathcal H)$ equipped with the Hilbert--Schmidt inner product $\langle X,Y\rangle=\operatorname{Tr}(X^\dagger Y)$. The infinite-dimensional case admits an analogous treatment when a Liouvillian spectral gap exists. The system is coupled to two isomorphic reservoirs through two symmetric ports. Its evolution is governed by the Lindblad master equation
\begin{equation}\label{eq:Lindblad}\dot\rho = \mathcal L_{\eta,\lambda}[\rho] = -i[H_S,\rho] + \sum_{\nu=L,R}\mathcal D_{P_\nu}[\rho],\end{equation}
where $\rho$ is the reduced density matrix, $H_S$ is the system Hamiltonian, $\mathcal D_{P_\nu}$ is the dissipator for reservoir $\nu=L,R$, with $P_\nu$ denoting the set of reservoir parameters. We introduce a dimensionless perturbation strength $\lambda\ll 1$ and a direction parameter $\eta\in(-1,1)$, and set
\begin{equation}\label{eq:PLPR}P_L = P_0 + \eta\lambda P_1,\qquad P_R = P_0 - \eta\lambda P_1,\end{equation}
so that the two reservoirs are biased equally but oppositely by $\pm\eta\lambda P_1$ about the common reference $P_0$. The operation $\eta\to-\eta$ reverses the sign of the bias, swapping the roles of the two reservoirs. We call this the swap, a discrete operation realized in practice by interchanging the reservoir parameters, as illustrated in Fig.~\ref{fig:1}. The Liouvillian $\mathcal L_{\eta,\lambda}$ can then be expanded analytically about $\eta\lambda=0$:
\begin{equation}\mathcal L_{\eta,\lambda} = \mathcal L^{(0)} + \eta\lambda \mathcal L^{(1)} + (\eta\lambda)^2 \mathcal L^{(2)} + \cdots,\end{equation}
where $\mathcal L^{(n)}$ is independent of $\eta\lambda$. 

The leading term $\mathcal L^{(0)}=\mathcal L_{\eta,\lambda}|_{\eta\lambda=0}$ describes the reference system with two identical reservoirs. It is a GKSL generator, trace-preserving and completely positive, with a unique steady state $\rho_{\text{ss}}^{(0)}$ and all other eigenvalues strictly negative in real part. Throughout this paper, \emph{decay rate} refers to the real part of a Liouvillian eigenvalue, i.e. the rate at which the corresponding eigenmode decays. Let $\operatorname{spec}(\mathcal L^{(0)})=\{\omega_0^{(0)}=0,\omega_1^{(0)},\omega_2^{(0)},\dots\}$ be ordered by 
\begin{equation}0 > \operatorname{Re}\omega_1^{(0)} \ge \operatorname{Re}\omega_2^{(0)} > \operatorname{Re}\omega_3^{(0)} \ge \cdots,\end{equation}
with $\omega_1^{(0)}$ and $\omega_2^{(0)}$ assumed non-degenerate. Define the slow subspace as the span of the eigenoperators of $\omega_1^{(0)}$ and $\omega_2^{(0)}$, with left and right eigenoperators satisfying $\operatorname{Tr}[(l_i^{(0)})^\dagger r_j^{(0)}]=\delta_{ij}$ ($i,j=1,2$). The fast subspace ($\omega_3^{(0)}$, $\omega_4^{(0)}$,...) need not be diagonalizable. The slow and fast subspaces are separated by a spectral gap $\Delta=\operatorname{Re}\omega_2^{(0)}-\operatorname{Re}\omega_3^{(0)}>0$.

\textit{Slow-mode reduction}. -- We investigate the two slowest relaxation modes of $\mathcal L_{\eta,\lambda}$. Focusing on the two slowest modes is common in the open QME literature. Since $\mathcal L_{\eta,\lambda}$ is a small perturbation of $\mathcal L^{(0)}$ by the bias, we obtain the effective dynamics of $\mathcal L_{\eta,\lambda}$'s two slowest modes via perturbation theory based on the slow subspace of $\mathcal L^{(0)}$. The Riesz spectral projection gives~\cite{Kato1995}
\begin{equation}P_\sigma = \frac{1}{2\pi i}\oint_{\Gamma_\sigma}(z-\mathcal L^{(0)})^{-1}\,dz,\end{equation}
where $\Gamma_\sigma$ is a positively oriented contour enclosing the isolated spectral set $\sigma$. The projection satisfies $P_\sigma^2=P_\sigma$, $[\mathcal L^{(0)},P_\sigma]=0$, $P_\sigma P_\tau=0$ for $\sigma\cap\tau=\varnothing$, and summing over the spectral partition gives $\sum_i P_{\sigma_i}=\mathbf 1$. Let $\Pi$ denote the projection onto the slow subspace and $\Pi_{\text{rest}}=\mathbf 1-\Pi$. Applying the Feshbach projection method~\cite{V2014PRA} to first order in the perturbation yields an effective $2\times2$ generator (see the Supplemental Material):
\begin{equation}\label{eq:Leff}\mathcal L_{\text{eff}} = \begin{pmatrix} \omega_1^{(0)} + \eta\lambda (\mathcal L^{(1)})_{11} & \eta\lambda (\mathcal L^{(1)})_{12} \\\eta\lambda (\mathcal L^{(1)})_{21} & \omega_2^{(0)} + \eta\lambda (\mathcal L^{(1)})_{22} \end{pmatrix},\end{equation}
where $(\mathcal L^{(1)})_{ab}=\operatorname{Tr}[(l_a^{(0)})^\dagger \mathcal L^{(1)} r_b^{(0)}]$ ($a,b=1,2$). Corrections from the fast subspace enter at $O((\eta\lambda)^2/\Delta)$, and those beyond first-order perturbation at $O((\eta\lambda)^2)$. The spectral gap $\Delta$ must be substantial, ensuring a clear separation between the slow and fast subspaces. Within this regime the two slowest relaxation modes of $\mathcal L_{\eta,\lambda}$ are governed by $\mathcal L_{\text{eff}}$, expressed in the Liouville-space basis of $\mathcal L^{(0)}$'s slow eigenoperators.

Swapping the reservoirs is equivalent to transforming the system while the reservoirs stay fixed. Owing to the structural symmetry, this transformation can be realized by a unitary $\mathcal S$ on the system Hilbert space satisfying $\mathcal S \mathcal L_{\eta,\lambda} \mathcal S^{-1} = \mathcal L_{-\eta,\lambda}$, a Liouvillian operation within the Kawabata classification~\cite{Kawabata2023PRXQ,B2012NJP}. The operation $\mathcal S$ acts on the system, not on the reservoir parameter $\eta$. This unitary transformation preserves the spectrum exactly, so $\mathcal L_{\text{eff}}(+\eta)$ and $\mathcal L_{\text{eff}}(-\eta)$ share every eigenvalue, spectral gap, and Jordan-block pattern. The steady state is unique for each bias direction, and the two are mirror images: $\rho_{\text{ss}}(-\eta)=\mathcal S\rho_{\text{ss}}(\eta)\mathcal S^{-1}$. Any difference between the two bias directions in the relaxation of a fixed initial state must therefore come from the eigenvectors of $\mathcal L_{\text{eff}}$, which rotate under $\mathcal S$.

\textit{Criterion for NQME}. -- The unitary $\mathcal S$ enforces $(\mathcal L^{(1)})_{11}=(\mathcal L^{(1)})_{22}=0$ (Supplemental Material) in Eq.~\eqref{eq:Leff}, so the eigenvalues of $\mathcal L_{\text{eff}}$ are even in $\eta$. They are given by \begin{equation}\label{eq:eigenvalues}\mu_\pm = s \pm \sqrt{\mathcal A},\end{equation} where $u = \eta\lambda(\mathcal L^{(1)})_{12}$, $v = \eta\lambda(\mathcal L^{(1)})_{21}$, $s = \tfrac12(\omega_1^{(0)}+\omega_2^{(0)})$, $\delta = \tfrac12(\omega_1^{(0)}-\omega_2^{(0)})$, and $\mathcal A = \delta^2 + uv$. The right and left eigenvectors are
\begin{equation}
\begin{aligned}
v_\pm &= \frac{1}{\sqrt{N_\pm}}\bigl(u,\; -\delta\pm\sqrt{\mathcal A}\bigr)^T,\\
w_\pm &= \frac{1}{\sqrt{N_\pm}}\bigl(v,\; -\delta\pm\sqrt{\mathcal A}\bigr),
\end{aligned}
\label{eq:eigenvectors}
\end{equation}
satisfying $w_i\cdot v_j=\delta_{ij}$, with $N_\pm=2\sqrt{\mathcal A}(\sqrt{\mathcal A}\mp\delta)$.

Let $\delta\rho(t)=\rho(t)-\rho_{\text{ss}}(\eta)$. Its slow component is
\begin{equation}\label{eq:deltarho}\delta\rho_{\text{slow}}(t)\simeq c_+ e^{\mu_+ t}v_+ + c_- e^{\mu_- t}v_-,\end{equation}
Here $\mu_+$ and $\mu_-$ determine the decay rates, and $c_\pm = w_\pm \cdot a$, with $a$ representing the initial-state coordinate in $L^{(0)}$'s slow subspace, defined as
\begin{equation}
a = (\operatorname{Tr}[(l_1^{(0)})^\dagger\rho(0)],\operatorname{Tr}[(l_2^{(0)})^\dagger\rho(0)])^T.
\end{equation}
The steady state of $\mathcal L_{\text{eff}}$ is orthogonal to its non-steady eigenmodes, so its coordinates contribute nothing to $c_\pm$, consistent with the fact that the steady state does not relax.

To determine whether the initial state decouples from the slowest eigenmode, we introduce the decoupling vector
\begin{equation}\label{eq:chi}\chi(\eta,\lambda)=\big(v,-\delta+\sqrt{\mathcal A}\big).\end{equation}
The decoupling condition is $\chi\cdot a=0$. Any state closer to the steady state than the decoupled one necessarily has a nonzero projection on the slowest eigenmode. Such closer states relax at the slowest rate, while the decoupled state bypasses the slowest mode, indicating the strong QME.

\emph{Definition (NQME). Let $\rho(0)$ be a fixed initial state with coordinates $a$. If there exists $\eta$ such that $\chi(\eta)\cdot a=0$ and $\chi(-\eta)\cdot a\neq0$, then the state exhibits the NQME at $\eta$}

We refer to the bias direction that yields decoupling as the forward direction. The opposite direction, corresponding to swapping the two reservoirs ($\eta\to-\eta$), is the reverse direction. The initial state can be prepared independently of the reservoir parameters (e.g., by a microwave pulse or a global radio-frequency pulse on the qubits), so the definition is directly realizable. The criterion $\chi(\eta)\cdot a=0$ is an algebraic condition on the projection of the initial state onto the slowest mode, independent of any distance measure.

\emph{Condition L1 (direction-dependent decoupling). The off-diagonal elements of $\mathcal L_{\text{eff}}$ are nonzero.}

The condition~L1 is a necessary condition for NQME. It originates from the requirement that $\chi(+\eta)$ and $\chi(-\eta)$ are linearly independent and the eigenvectors of $\mathcal L_{\text{eff}}$ are nonzero. Equivalently, L1 means that the bias genuinely couples the two different slow modes of $\mathcal L^{(0)}$, so that the resulting eigenvectors of $\mathcal L_{\text{eff}}$ become $\eta$-dependent and rotate under the swap.

L1 guarantees $\eta$-dependent decoupling under the swap. Realizing this decoupling, however, also demands the right initial state. The vectors $\chi(\eta)$ and $\chi(-\eta)$ are related by the unitary $\mathcal S$. A unitary preserves inner products, so if $\rho(0)$ decouples from $\chi(\eta)$, its $\mathcal S$-transform $\mathcal S\rho(0)\mathcal S^{-1}$ decouples from $\chi(-\eta)$. Consequently, when $\rho(0)$ and $\eta$ meet the NQME definition, $\mathcal S\rho(0)\mathcal S^{-1}$ produces QME in the reverse direction and none in the forward. If $\rho(0)$ were $\mathcal S$-invariant, unitarity would force it to decouple from $\chi(-\eta)$ whenever it decouples from $\chi(\eta)$. Because $\chi(\eta)$ and $\chi(-\eta)$ are linearly independent under L1, a nonzero state in the slow subspace cannot be orthogonal to both. The NQME therefore requires $\rho(0)\neq\mathcal S\rho(0)\mathcal S^{-1}$.

The relaxation trajectory $\rho(t)$ is determined by two independent elements -- the dynamics governed by $\mathcal L_{\eta,\lambda}$, and the initial condition $\rho(0)$. If $\mathcal S$ were applied to both, the entire trajectory under $+\eta$ would be the $\mathcal S$-image of the trajectory under $-\eta$, and the NQME would vanish. The nonreciprocal Mpemba effect originates from applying $\mathcal S$ to $\mathcal L_{\eta,\lambda}$ while withholding it from $\rho(0)$.

This exposes a more general structure. The swap acts on every element of the dynamics -- $\mathcal L_{\eta,\lambda}$ and all it determines. The initial state, by contrast, is exempted by the NQME definition itself, which requires a fixed $\rho(0)$. Because of this asymmetry, the eigenvectors rotate under the swap while $\rho(0)$ stays fixed. Their overlap changes -- what is orthogonal in one bias direction becomes non-orthogonal in the other -- and the Mpemba effect turns on or off accordingly.

For a fixed initial state, the two directions differ not in how fast the eigenmodes relax, but in which eigenmode governs the relaxation. The pinned spectrum rules out a spurious QME. If the two bias directions had independent spectra, the decoupled state could appear to overtake the near state in the forward direction while still relaxing more slowly than the reverse direction's slowest mode. This is a false acceleration. Spectral pinning prevents it, because both directions share the same decay rates. The ratio $R=|\mathcal Re(\mu_-)/\mathcal Re(\mu_+)|>1$ confirms that the decoupled state decays faster in the forward direction than it does in the reverse direction, and faster than the near state in both directions. The speedup is genuine.

The NQME persists at EP, where the mechanism takes a different form. At $\mathcal A=0$, the two eigenvalues merge and the eigenvectors coalesce. The ratio $R$ therefore no longer applies. Instead, $\mathcal L_{\text{eff}}$ is non-diagonalizable, reducing to the Jordan-block form $\mathcal L_{\text{eff}}=sI+N(\eta)$ with $N(\eta)^2=0$. The propagator $e^{\mathcal L_{\text{eff}}t}=(I+N(\eta)t)e^{st}$ implies critical slowing down with the form $t e^{st}$. For the coordinates $a$ of an initial state, decoupling ($\chi(\eta)\cdot a=0$) yields $N(\eta)a=0$, cancelling the critical slowing so the state relaxes as a pure exponential $e^{st}$. In the opposite direction $\chi(-\eta)\cdot a\neq0$, the critical slowing survives. The NQME at an EP is therefore a nonreciprocal critical slowing. The direction selectively protects the far state from the slowing. The crucial point is that at an EP the two decay rates merge into one. Yet the directional asymmetry persists, carried by the Jordan (generalized eigenvector) structure. This is the spectrum-independent effect in its purest form.

\textit{Example: A three-level system coupled to two Fermi reservoirs}. -- We take a three-level system ${|0\rangle,|L\rangle,|R\rangle}$, where $|L\rangle,|R\rangle$ are degenerate excited states. The reservoir parameters are $P_\nu={\gamma_\nu,f_\nu,\gamma_\phi^\nu}$ ($\nu=L,R$). We set $\gamma_\nu=\gamma_0+s_\nu\eta\lambda b_\gamma$, $f_\nu=f_0+s_\nu\eta\lambda b_f$, $\gamma_\phi^\nu=\gamma_\phi+s_\nu\eta\lambda b_\phi$, with $s_L=-s_R=1$, where $\gamma_\nu$ is the decay rate, $f_\nu$ the Fermi occupation, and $\gamma_\phi^\nu$ the dephasing rate. The system operation $\mathcal S$, equivalent to directly interchanging the reservoir parameters, is realized by swapping $|L\rangle$ and $|R\rangle$. For $\gamma_\phi>\gamma_0(1+f_0)$, the two slowest modes of $\mathcal L^{(0)}$ contain no coherence terms, and (see the Supplemental Material)
\begin{equation}\label{eq:LeffModel}\mathcal L_{\text{eff}}=\begin{pmatrix} -\gamma_0(1-f_0) & \eta\lambda[\gamma_0 b_f+(1+f_0)b_\gamma] \\ -\eta\lambda[\gamma_0 b_f-(1-f_0)b_\gamma] & -\gamma_0(1+f_0) \end{pmatrix},\end{equation}
A direct computation gives $(\mathcal L^{(1)})_{11}=(\mathcal L^{(1)})_{22}=0$, so the diagonal of $\mathcal L_{\text{eff}}$ is independent of $\eta$. Hence the eigenvalues do not change under $\eta\to-\eta$, verifying that the spectrum is frozen.

The spectral gap is $\Delta=\tfrac12[\gamma_\phi-\gamma_0(1+f_0)]$, tuned by dephasing. A strong dephasing rate ensures a sufficiently large spectral gap. The dephasing rate does not enter $\mathcal L_{\text{eff}}$. Its presence allows us to obtain a compact analytic form, but is not necessary for the mechanism. Without dephasing, the slowest subspace of $\mathcal L^{(0)}$ would contain coherence terms, complicating the spectral gap and making an analytic treatment considerably more involved. The reservoir-induced excitation rate $|0\rangle\!\to\!|\nu\rangle$ is $\gamma_\nu f_\nu$, and the relaxation rate $|\nu\rangle\!\to\!|0\rangle$ is $\gamma_\nu(1-f_\nu)$. A nonzero $b_\gamma$ or $b_f$ reverses the rates and swaps the roles of the two reservoirs under $\eta\to-\eta$. We therefore set $b_\gamma=0$, giving
\begin{equation}\label{lef2}
\mathcal L_{\text{eff}}=-\gamma_0 I+\gamma_0 f_0\sigma_z+ i\eta\lambda\gamma_0 b_f\sigma_y.\end{equation}
Then $u=\eta\lambda\gamma_0 b_f$, $v=-\eta\lambda\gamma_0 b_f$, $\delta=\gamma_0 f_0$, $\mathcal A=\gamma_0^2(f_0^2-\eta^2\lambda^2 b_f^2)$, and the decoupling vector is $\chi(\eta)=\gamma_0(-\eta\lambda b_f,g-f_0)$ with $g=\sqrt{f_Lf_R}$. The vectors $\chi(+\eta)$ and $\chi(-\eta)$ are linearly independent and nonzero -- condition L1 is strictly satisfied. Without the coherence terms $\mathcal S\rho(0)\mathcal S^{-1}$ merely exchanges the populations of $|L\rangle$ and $|R\rangle$. The NQME therefore requires an explicit L--R population imbalance, since an $\mathcal S$-invariant $\rho(0)$ with equal populations cannot satisfy the definition.

\begin{figure}[t]
\centering
\includegraphics[width=\columnwidth]{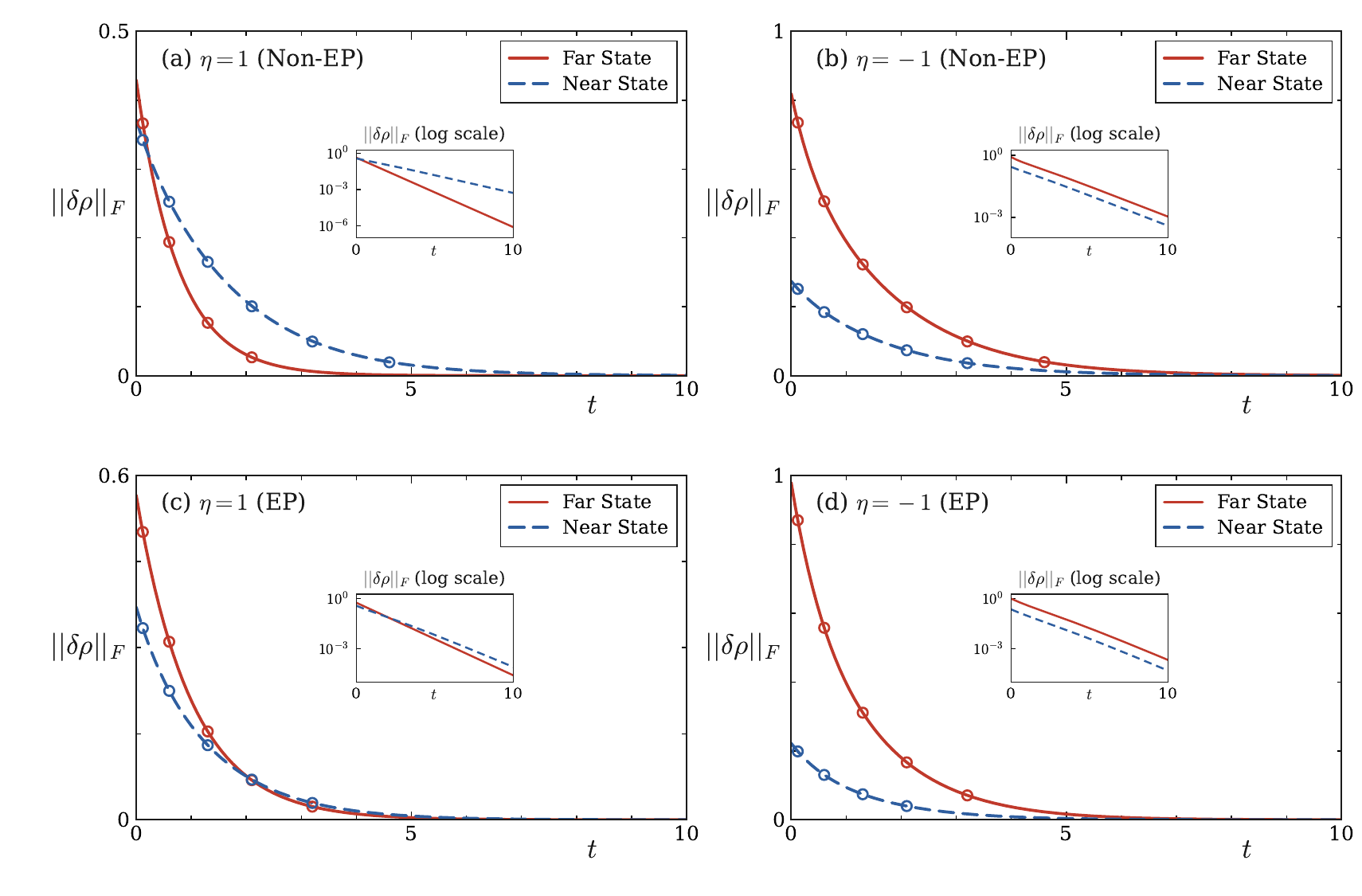}
\caption{(color online) The time evolution of the Frobenius-norm distance to the steady state. Solid and dashed lines show the evolution computed from the $2\times2$ effective model $\mathcal L_{\text{eff}}$. Open circles show the evolution from the full $9\times9$ superoperator $\mathcal L_{\eta,\lambda}$ with $\gamma_\phi=5$. Each panel includes a logarithmic-scale inset. (a) and (b) show the forward direction ($\eta=1$) and reversed direction ($\eta=-1$), respectively, in the non-EP regime. (c) and (d) show the forward and reversed directions, respectively, at the EP. Parameters are $\lambda=0. 2$, $b_f=\gamma_0=1$, $f_0=0.38$ for non-EP regime, $f_0=0.2$ at EP. Initial states, represented in the ${|0\rangle,|L\rangle,|R\rangle}$ basis, are $\rho_{\text{far}}=\operatorname{diag}(0.0385,0.7533,0.2082)$, $\rho_{\text{near}}=\operatorname{diag}(0.33,0.30,0.37)$ for non-EP regime. $\rho_{\text{far}}=\operatorname{diag}(0.2,0.8,0)$, $\rho_{\text{near}}=\operatorname{diag}(0.53,0.18,0.29)$ at EP.}
\label{fig:2}
\end{figure}

Figure~\ref{fig:2} confirms the NQME in both the non-EP and EP regimes. The agreement between the $2\times2$ effective model and the full $9\times9$ Liouvillian is shown.

\textit{Discussion}. -- We have demonstrated a nonreciprocal quantum Mpemba effect. Swapping the two reservoirs turns the Mpemba effect on and off for a fixed initial state. The mechanism separates the Liouvillian eigenvectors from the eigenvalue spectrum---the swap leaves the spectrum intact and rotates only the eigenvectors, the channels through which the system relaxes. This separation follows from a structural symmetry of the two-port Liouvillian and applies to any open system with two symmetric ports coupled to biased reservoirs.
	
Reservoir engineering is by now a mature experimental capability~\cite{Harrington2022}. In quantum-dot setups, a temperature bias between the electronic leads can be applied and tuned via local heating~\cite{Josefsson2018}. Analogous in-situ control has been demonstrated for engineered bosonic reservoirs in trapped ions~\cite{So2024SciAdv,So2025arXiv} and superconducting circuits~\cite{Tan2017NatCommun,Murch2012}. The coupling to and dissipation induced by the engineered reservoir can be tuned~\cite{So2024SciAdv,Tan2017NatCommun,Murch2012}, and its effective temperature is tunable as well~\cite{So2025arXiv,Tan2017NatCommun,Murch2012}.

A bias between the two reservoirs creates a generalized thermodynamic force, or affinity, between the reservoirs~\cite{Onsager1931PRL,Esposito2009RMP,Landi2022RMP}. Swapping L and R reverses this force. The NQME embodies a class in which the global behavior changes~\cite{Zhong2025npj} under the reverse. When L and R exchange roles, the eigenvectors rotate and the Mpemba character flips. 
\acknowledgments
This work was supported by the Natural Science Foundation of Shandong Province (Grants No.~ZR2025LLZ004 and ZR2024LLZ002), the NSFC (Grants No.~11505023, 11974209, 12274257, and U25A6009), the MOST project (Grant No. 2025YFE0217600), and the QNMP (Grant No. 2021ZD0301800).

\end{document}


\begin{center}
{\Large \textbf{Supplementary Material: Nonreciprocal Mpemba Effect}}
\end{center}

\bigskip

\section*{Supplementary Material}

This Supplementary Material provides the derivation of the slow-subspace reduction technique and the calculations for the three-level Fermi-reservoir model presented in the main text.

\section{Slow-subspace reduction technique}
\subsection{Derivation of $\mathcal{L}_{\mathrm{eff}}$}

Let $\Pi$ be the spectral projection onto the slow subspace (corresponding to eigenvalues $\omega_1^{(0)}$ and $\omega_2^{(0)}$ of $\mathcal{L}^{(0)}$), and $\Pi_{\mathrm{rest}} \equiv \mathbf{1} - \Pi$. Partitioning the Liouville superoperator $\mathcal{L}_{\eta,\lambda}$ accordingly yields
\begin{equation}
\mathcal{L}_{\eta,\lambda} = A+B+C+D,
\label{eq:S1}
\end{equation}
with the blocks $A \equiv \Pi \mathcal{L}_{\eta,\lambda}\Pi$, $B \equiv \Pi \mathcal{L}_{\eta,\lambda}\Pi_{\mathrm{rest}}$, $C \equiv \Pi_{\mathrm{rest}} \mathcal{L}_{\eta,\lambda}\Pi$, and $D \equiv \Pi_{\mathrm{rest}} \mathcal{L}_{\eta,\lambda}\Pi_{\mathrm{rest}}$.

Left-multiplying the eigenvalue equation $\mathcal{L}_{\eta,\lambda} \psi= \omega \psi$ by $\Pi$ and $\Pi_{\mathrm{rest}}$ gives
\begin{align}
A \psi_\Pi + B \psi_{\mathrm{rest}} &= \omega \psi_\Pi,
\label{eq:S2} \\
C \psi_\Pi + D \psi_{\mathrm{rest}} &= \omega \psi_{\mathrm{rest}},
\label{eq:S3}
\end{align}
with $\psi_\Pi = \Pi \psi$ and $\psi_{\mathrm{rest}} = \Pi_{\mathrm{rest}} \psi$, so that $\psi = \psi_\Pi + \psi_{\mathrm{rest}}$. For eigenvalues in the slow sector, $\omega \notin \operatorname{spec}(D)$ is guaranteed by the spectral gap $\Delta$. From \eqref{eq:S3}, $\psi_{\mathrm{rest}} = (\omega - D)^{-1} C \psi_\Pi$. Substituting into \eqref{eq:S2} yields the Feshbach-reduced equation~[1]:
\begin{equation}
\mathcal{L}_{\mathrm{eff}}(\omega)\, \psi_\Pi = \omega \psi_\Pi,
\label{eq:S4a}
\end{equation}
where
\begin{equation}
\mathcal{L}_{\mathrm{eff}}(\omega) \equiv A + B(\omega - D)^{-1}C.
\label{eq:S4b}
\end{equation}

Let
\begin{equation}
V \equiv \mathcal{L}_{\eta,\lambda} - \mathcal{L}^{(0)} = \eta\lambda \mathcal{L}^{(1)} + \frac{(\eta\lambda)^2}{2!} \mathcal{L}^{(2)} + \cdots.
\label{eq:S5}
\end{equation}
Since $\Pi \mathcal{L}^{(0)}\Pi_{\mathrm{rest}} = \Pi_{\mathrm{rest}} \mathcal{L}^{(0)}\Pi = 0$, the blocks are
\begin{align}
A &= \Pi \mathcal{L}^{(0)}\Pi + \Pi V\Pi =
\begin{pmatrix}
\omega_1^{(0)} & 0 \\
0 & \omega_2^{(0)}
\end{pmatrix}
+ \Pi V\Pi,
\label{eq:S6} \\
B &= \Pi V \Pi_{\mathrm{rest}}, \notag \\
C &= \Pi_{\mathrm{rest}} V \Pi,
\label{eq:S7} \\
D &= \Pi_{\mathrm{rest}} \mathcal{L}^{(0)}\Pi_{\mathrm{rest}} + \Pi_{\mathrm{rest}} V \Pi_{\mathrm{rest}}
\equiv D_0 + W,
\quad D_0 \equiv \Pi_{\mathrm{rest}} \mathcal{L}^{(0)}\Pi_{\mathrm{rest}}.
\label{eq:S8}
\end{align}
Expanding $(\omega - D)^{-1}$ as a Neumann series,
\begin{equation}
(\omega - D)^{-1} = (\omega - D_0)^{-1} + (\omega - D_0)^{-1} W (\omega - D_0)^{-1} + \cdots.
\label{eq:S9}
\end{equation}
Substituting \eqref{eq:S5}--\eqref{eq:S9} into \eqref{eq:S4b} yields
\begin{equation}
\begin{aligned}
\mathcal{L}_{\mathrm{eff}} =
\begin{pmatrix}
\omega_1^{(0)} + \eta\lambda (\mathcal{L}^{(1)})_{11} & \eta\lambda (\mathcal{L}^{(1)})_{12} \\
\eta\lambda (\mathcal{L}^{(1)})_{21} & \omega_2^{(0)} + \eta\lambda (\mathcal{L}^{(1)})_{22}
\end{pmatrix}
\\
+ (\eta\lambda)^2 \sum_{c \notin \{1,2\}}
\frac{(\mathcal{L}^{(1)})_{ac}(\mathcal{L}^{(1)})_{cb}}{\omega - \omega_c^{(0)}}
+ O((\eta\lambda)^3),
\end{aligned}
\label{eq:S10}
\end{equation}
where we define the matrix elements
\begin{equation}
(\mathcal{L}^{(1)})_{ab} \equiv \operatorname{Tr}[(l_a^{(0)})^\dagger \mathcal{L}^{(1)} r_b^{(0)}], \quad a,b \in \{1,2\}.
\label{eq:S11}
\end{equation}
and have truncated $V$ to first order. The sum over $c$ in \eqref{eq:S10} includes the steady-state index $0$ and all fast-mode indices $k \ge 3$. The contribution from $c=0$ vanishes because the steady-state left eigenvector is the identity, as a consequence of trace preservation. The denominators are controlled by the spectral gap $\Delta=\omega_2^{(0)} - \omega_3^{(0)}$, rendering the sum $O((\eta\lambda)^2/\Delta)$. Hence
\begin{equation}
\mathcal{L}_{\mathrm{eff}} =
\begin{pmatrix}
\omega_1^{(0)} + \eta\lambda (\mathcal{L}^{(1)})_{11} & \eta\lambda (\mathcal{L}^{(1)})_{12} \\
\eta\lambda (\mathcal{L}^{(1)})_{21} & \omega_2^{(0)} + \eta\lambda (\mathcal{L}^{(1)})_{22}
\end{pmatrix}
+ O\!\left(\frac{(\eta \lambda)^2}{\Delta}\right) + O((\eta\lambda)^2).
\label{eq:S12}
\end{equation}
The second error term originates from truncating $V$ to first order in $\eta\lambda$.

\section{Unitary superoperator $\mathcal S$}

We assume that there exists a unitary superoperator $\mathcal S$ obeying
\begin{equation}
\mathcal S \mathcal{L}(\eta,\lambda) \mathcal S^{-1} = \mathcal{L}(-\eta,\lambda).
\label{eq:S13}
\end{equation}
$\mathcal S$ acts on operators and state vectors in the system Hilbert space but not on the scalar parameter $\eta$. Substituting Eq.~(3) in the main text into \eqref{eq:S13} yields
\begin{align}
\mathcal S \mathcal{L}^{(0)} \mathcal S^{-1} &= \mathcal{L}^{(0)},
\label{eq:S14a} \\
\mathcal S \mathcal{L}^{(1)} \mathcal S^{-1} &= - \mathcal{L}^{(1)}.
\label{eq:S14b}
\end{align}
Thus $[\mathcal{L}^{(0)},\mathcal S]=0$ and $\{\mathcal{L}^{(1)},\mathcal S\}=0$. Consequently, $r_1^{(0)}$ and $r_2^{(0)}$ satisfy
\begin{equation}
\mathcal S r_a^{(0)} \mathcal S^{-1} = p_a r_a^{(0)},\qquad p_a \in \{+1, -1\},\quad a=1,2.
\label{eq:S15}
\end{equation}
Biorthonormality $\operatorname{Tr}[(l_a^{(0)})^\dagger r_b^{(0)}] = \delta_{ab}$ and unitarity of $\mathcal S$ imply the same parity for the left vectors:
\begin{equation}
\mathcal S l_a^{(0)} \mathcal S^{-1} = p_a l_a^{(0)}.
\label{eq:S16}
\end{equation}
Using the invariance of the trace under unitary transformations, we obtain
\begin{equation}
\begin{aligned}
(\mathcal{L}^{(1)})_{aa}
&= \operatorname{Tr}\big[(\mathcal S l_a^{(0)} \mathcal S^{-1})^\dagger\,
(\mathcal S \mathcal{L}^{(1)} \mathcal S^{-1})\,
(\mathcal S r_a^{(0)} \mathcal S^{-1})\big] \\
&= -p_a^2(\mathcal{L}^{(1)})_{aa}= - (\mathcal{L}^{(1)})_{aa}.
\end{aligned}
\label{eq:S17}
\end{equation}
Therefore $(\mathcal{L}^{(1)})_{11} = (\mathcal{L}^{(1)})_{22} = 0$.

\section{Three-level Fermi-reservoir model}
\subsection{Hamiltonian and bath parametrization}

The system Hilbert space is spanned by $\{|0\rangle, |L\rangle, |R\rangle\}$, with $|L\rangle$ and $|R\rangle$ as degenerate excited states ($\omega_L = \omega_R = \omega$). The system Hamiltonian is $H_S = \operatorname{diag}(0, \omega, \omega)$. The left and right thermal reservoirs couple to the transitions $|L\rangle\leftrightarrow|0\rangle$ and $|R\rangle\leftrightarrow|0\rangle$, respectively. Each reservoir is characterized by three parameters: the decay rate $\gamma_\nu$, the Fermi occupation factor $f_\nu$, and the pure dephasing rate $\gamma_\phi^\nu$ ($\nu = L, R$). These are parametrized as
\begin{align}
\gamma_\nu &= \gamma_0 + s_\nu \eta\lambda\, b_\gamma,\notag\\
f_\nu &= f_0 + s_\nu \eta\lambda\, b_f,\notag\\
\gamma_\phi^\nu &= \gamma_\phi + s_\nu \eta\lambda\, b_\phi,\notag
\end{align}
with $s_L = +1$ and $s_R = -1$.

The Lindblad master equation reads
\begin{equation}
\dot\rho = -i[H_S, \rho] + \sum_{\nu=L,R} \mathcal D_{P_\nu}[\rho],
\label{eq:S18}
\end{equation}
where $P_\nu = \{\gamma_\nu, f_\nu, \gamma_\phi^\nu\}$ and $\mathcal D_{P_\nu}[\rho] \equiv \gamma_\nu f_\nu\, \mathcal D_J[|\nu\rangle\langle 0|]\rho + \gamma_\nu(1-f_\nu)\, \mathcal D_J[|0\rangle\langle \nu|]\rho + \gamma_\phi^\nu\, \mathcal D_J[|\nu\rangle\langle \nu|]\rho$, with $\mathcal D_J[J]\rho \equiv J\rho J^\dagger - \tfrac12\{J^\dagger J, \rho\}$.

\subsection{Closed population subspace}

Let $\mathbf p = (\rho_{00}, \rho_{LL}, \rho_{RR})^{\!T}$ denote the population vector. From \eqref{eq:S18}, we obtain the closed dynamical equation $\dot{\mathbf p} = M \mathbf p$, with
\begin{equation}
M = \begin{pmatrix}
-(\gamma_L f_L + \gamma_R f_R) & \gamma_L(1-f_L) & \gamma_R(1-f_R) \\
\gamma_L f_L & -\gamma_L(1-f_L) & 0 \\
\gamma_R f_R & 0 & -\gamma_R(1-f_R)
\end{pmatrix}.
\label{eq:S19}
\end{equation}
The three coherences $\rho_{0L}$, $\rho_{0R}$, $\rho_{LR}$ and their complex conjugates evolve independently. Their decay rates at $\lambda=0$ are given by
\begin{align}
\Gamma_{0L}^{(0)} &= \Gamma_{0R}^{(0)}
= \tfrac12\big[\gamma_0(1+f_0) + \gamma_\phi\big],
\label{eq:S20a} \\
\Gamma_{LR}^{(0)} &= \gamma_0(1-f_0) + \gamma_\phi.
\label{eq:S20b}
\end{align}

\subsection{Zeroth-order Liouvillian spectrum and eigenvectors}

At $\lambda=0$, the population block reduces to
\begin{equation}
M^{(0)} = \gamma_0
\begin{pmatrix}
-2f_0 & 1-f_0 & 1-f_0 \\
f_0 & -(1-f_0) & 0 \\
f_0 & 0 & -(1-f_0)
\end{pmatrix}.
\label{eq:S21}
\end{equation}
Each column of $M^{(0)}$ sums to zero, yielding one eigenvalue of zero. The other two nonzero eigenvalues are $-\gamma_0(1-f_0)$ and $-\gamma_0(1+f_0)$.

Under the condition $\gamma_\phi > \gamma_0(1+f_0)$, the coherence decay rates in \eqref{eq:S20a} are significantly larger than the magnitudes of the two nonzero eigenvalues of $M^{(0)}$. Consequently, the steady mode and the two slowest modes of $\mathcal{L}^{(0)}$ reside entirely in the population subspace described by \eqref{eq:S21}.

The right and left steady-state eigenvectors of $\mathcal{L}^{(0)}$ are
\begin{align}
r_0^{(0)} &= \frac{1}{1+f_0}(1-f_0,\; f_0,\; f_0)^T,
\label{eq:S22a} \\
l_0^{(0)} &= (1,1,1).
\label{eq:S22b}
\end{align}
The right eigenvectors for the two nonzero slow modes of $\mathcal{L}^{(0)}$ are
\begin{align}
r_1^{(0)} &= (0,1,-1)^T,
\label{eq:S23a} \\
r_2^{(0)} &= (2,-1,-1)^T,
\label{eq:S23b}
\end{align}
with the corresponding left eigenvectors
\begin{align}
l_1^{(0)} &= (0,\tfrac12,-\tfrac12),
\label{eq:S24a} \\
l_2^{(0)} &= \frac{1}{2(1+f_0)}\big(2f_0,\; -(1-f_0),\; -(1-f_0)\big).
\label{eq:S24b}
\end{align}
In the operator representation,
\[
r_1^{(0)} = |L\rangle\langle L| - |R\rangle\langle R|,\qquad
r_2^{(0)} = 2|0\rangle\langle 0| - |L\rangle\langle L| - |R\rangle\langle R|.
\]

\subsection{First-order perturbation operator and its matrix elements}

Differentiating \eqref{eq:S18} with respect to $\eta\lambda$ and evaluating at $\eta\lambda = 0$ yields the first-order perturbation:
\begin{equation}
\begin{aligned}
\mathcal{L}^{(1)}\rho = &\;(\gamma_0 b_f + b_\gamma f_0)\,\mathcal D_J[|L\rangle\langle 0|]\rho
+ \big(b_\gamma(1-f_0) - \gamma_0 b_f\big)\,\mathcal D_J[|0\rangle\langle L|]\rho \\
&-\;(\gamma_0 b_f + b_\gamma f_0)\,\mathcal D_J[|R\rangle\langle 0|]\rho
- \big(b_\gamma(1-f_0) - \gamma_0 b_f\big)\,\mathcal D_J[|0\rangle\langle R|]\rho \\
&+\;b_\phi\,\mathcal D_J[|L\rangle\langle L|]\rho
- b_\phi\,\mathcal D_J[|R\rangle\langle R|]\rho .
\end{aligned}
\label{eq:S25}
\end{equation}
Applying $\mathcal{L}^{(1)}$ to $r_1^{(0)}$ and $r_2^{(0)}$ gives
\begin{align}
\mathcal{L}^{(1)} r_1^{(0)} &= \big[(1-f_0)b_\gamma - \gamma_0 b_f\big]\, r_2^{(0)},
\label{eq:S26} \\
\mathcal{L}^{(1)} r_2^{(0)} &= \big[(1+f_0)b_\gamma + \gamma_0 b_f\big]\, r_1^{(0)}.
\label{eq:S27}
\end{align}
Thus the matrix elements defined in \eqref{eq:S11} are
\begin{align}
(\mathcal{L}^{(1)})_{11} &= (\mathcal{L}^{(1)})_{22} = 0,
\label{eq:S28} \\
(\mathcal{L}^{(1)})_{12} &= \gamma_0 b_f + (1+f_0)b_\gamma,
\label{eq:S29a} \\
(\mathcal{L}^{(1)})_{21} &= -\gamma_0 b_f + (1-f_0)b_\gamma .
\label{eq:S29b}
\end{align}
Substituting \eqref{eq:S28}--\eqref{eq:S29b} into \eqref{eq:S12} yields the effective $2\times2$ generator presented in the main text:
\begin{equation}
\mathcal{L}_{\mathrm{eff}} =
\begin{pmatrix}
-\gamma_0(1-f_0) & \eta\lambda\big[\gamma_0 b_f + (1+f_0)b_\gamma\big] \\
-\eta\lambda\big[\gamma_0 b_f - (1-f_0)b_\gamma\big] & -\gamma_0(1+f_0)
\end{pmatrix}.
\label{eq:S30}
\end{equation}